# Large scale dynamics in two-dimensional turbulence

Zheng Ran

*Shanghai Institute of Applied Mathematics and Mechanics, Shanghai University, Shanghai 200072, P.R.China*

We consider freely decaying two-dimensional isotropic turbulence. It is usually assumed that, in such turbulence, the energy spectrum at small wave number $k$, takes the form $E(k \to 0) \propto I k^3$, where $I$ is the two-dimensional version of Loitsyansky's integral. In this paper, we developed a simple model for large scale dynamics of free decay two-dimensional turbulence based on the statistical solution of Navier-Stokes equation. We provide one possible explanation for the large scale dynamics in two-dimensional turbulence.

Two-dimensional turbulence has been a very active area of theoretical, numerical and experimental investigation, not only as an easier test case but also relevant to certain real quasi-two-dimensional situations. Examples including oceanic currents and atmospheric and geophysical flows, but two-dimensional flow are also realized in laboratory situations. However, the picture of two-dimensional turbulence remains not fully understood and in fact, there are some limitations to the classical KLB scenario. The mechanism of the 2D inverse cascade, however, is controversial, satisfying understanding of some important theoretical aspects of the two-dimensional turbulence, such as the large scale dynamics, the mechanism of the inverse cascade still call for substantial progress and upgrades [1-10].

In this paper, we are interested in the qualitative analysis about the physical mechanisms of the two-dimensional large scale dynamics in two-dimensional turbulence. By using the self-similar solution of two-dimensional Karman-Howarth equation, the energy spectrum function can be obtained by the integration of second-order velocity correlation function. Finally, we got the picture of the large scale dynamics which can be described by using the present theory straightly.

The most general form of the two-point one-time velocity correlation,

$$Q_{ij}(\vec{r},t) = \langle u_i(\vec{x},t) u_j(\vec{x}+\vec{r},t) \rangle = \langle u_i u_j' \rangle \qquad (1)$$

In incompressible isotropic two-dimensional turbulence is,

$$Q_{ij}(\vec{r},t) = u^2 \left\{ \frac{\partial}{\partial r}(rf)\delta_{ij} - \frac{r_i r_j}{r} f' \right\} \qquad (2)$$

Where $u^2 = \langle u_x^2 \rangle = \langle u_y^2 \rangle$, $r = |\vec{r}| = |\vec{x}' - \vec{x}|$, and $f(r,t)$ is the usual longitudinal correlation function. The triple velocity correlation tensor,

$$S_{ijk}(\vec{r},t) = \langle u_i(\vec{x},t) u_j(\vec{x},t) u_k(\vec{x}+\vec{r},t) \rangle = \langle u_i u_j u_k' \rangle \qquad (3)$$

On the other hand, takes the form,

$$S_{ijl} = u^3 \left\{ \frac{r_i \delta_{jl} + r_j \delta_{il}}{2r} \cdot \frac{\partial}{\partial r}(rK) - \frac{r_i r_j r_l}{r} \cdot \frac{\partial}{\partial r}\left(\frac{K}{r}\right) - \frac{r_l \delta_{ij}}{r} K \right\} \quad (4)$$

Where as usual, $u^3 K(r) = \langle u_x u_x u'_x \rangle$ (see, for example, Davidson, 2004).

As in three dimensions, the Navier-Stokes equation provides the evolution equation

$$\frac{\partial Q_{ij}}{\partial t} = \frac{\partial}{\partial r_l}\left[S_{jli} + S_{ilj}\right] + 2\nu \nabla^2 Q_{ij} \quad (5)$$

Setting $i = j$, after a little algebra we obtain the two-dimensional Karman-Howarth equation:

$$\frac{\partial}{\partial t}\left[u^2 r^3 f(r)\right] = u^3 \frac{\partial}{\partial r}\left[r^3 K\right] + 2\nu u^2 \frac{\partial}{\partial r}\left[r^3 f'(r)\right] \quad (6)$$

Let us suppose that the functions $f(r,t)$ and $K(r,t)$ preserve the same form as time increases with only the scale varying. Such functions will be termed as "self-preserving". Following Sedov [12,13], we introduce the new variables

$$\eta = \frac{r}{l(t)} \quad (7)$$

Where $l = l(t)$ is a uniquely specified similarity length scale. We will focus our attention on complete self-preserving solutions in the analysis below. For complete self-preserving isotropic turbulence, the Karman-Howarth equation takes the form

$$\frac{dK}{d\eta} + \frac{K}{\eta} = \frac{l}{2b^{\frac{3}{2}}} \cdot \frac{db}{dt} \cdot f - \frac{1}{2b^{\frac{1}{2}}} \cdot \frac{dl}{dt} \cdot \eta \frac{df}{d\eta} - \frac{\nu}{b^{\frac{1}{2}}} \cdot \frac{1}{l} \cdot \left(\frac{d^2 f}{d\eta^2} + \frac{3}{\eta} \cdot \frac{df}{d\eta}\right) \quad (8)$$

Analogy to the three dimensional turbulence case [14,15,16], one parameter family of exact solutions can be written in the form

$$f(\eta) = {}_1F_1\left(\sigma; 2; -\frac{a_1}{4}\eta^2\right) \quad (9)$$

Where $a_1$, $\sigma$ are the parameters introduced in Refs.[14,15,16]. The one-dimensional spectra that are most often measured are the one-dimensional Fourier transforms of a longitudinal or transverse correlation. We shall adopt the same convention as Tennekes and Lumley for the one-dimensional spectra of $u^2 f$, that is $F_{11}$. They are

$$F_{11}(k,t) = \frac{1}{\pi}\int_0^\infty u^2 f(r,t)\cos(kr)dr \quad (10)$$

With inverse transform,

$$u^2 f(r,t) = 2\int_0^\infty F_{11}(k,t)\cos(kr)dk \quad (11)$$

By using the integral formula:

$$\int_0^\infty {}_1F_1(a;c;-t^2)\cos(2zt)dt = \sqrt{\frac{\pi}{2}}\frac{\Gamma(c)}{\Gamma(a)}z^{2a-1}e^{-z^2}U\left(c-\frac{1}{2},a+\frac{1}{2},z^2\right) \quad (12)$$

Where ${}_1F_1(a,c,z)$ is Kummer's hypergeometric function, often written as $F(a,c,z)$, and $U(a,c,z)$ is another function closely related to Kummer's function defined by

$$U(a,c,z) = \frac{\Gamma(1-c)}{\Gamma(1+a-c)}F(a,c,z) + \frac{\Gamma(c-1)}{\Gamma(a)}z^{1-c}F(1+a-c,2-c,z) \quad (13)$$

Where $\Gamma(z)$ is the usual Gamma function.

Hence, we have

$$F_{11}(k,t) = \frac{2u^2}{\pi} \cdot \frac{l}{\sqrt{a_1}} \cdot \sqrt{\frac{\pi}{2}} \cdot \frac{\Gamma(2)}{\Gamma(\sigma)} \cdot \left(\frac{l}{\sqrt{a_1}}k\right)^{2\sigma-1} \cdot e^{-\frac{l^2}{a_1}k^2} \cdot U\left(\frac{3}{2},\sigma+\frac{1}{2},\frac{l^2}{a_1}k^2\right) \quad (14)$$

This is the exact result of the one-dimensional spectra based on the Sedov-type solution. The transformation equation between one- and n-dimensional spectra in the n-dimensional isotropic vector or scalar fluctuation field is derived by Hiroshi [11]. It is readily confirmed, that

$$E(k,t) = -\frac{k^3}{2} \cdot \int_k^\infty \frac{d}{dk_1}\left\{\frac{1}{k_1} \cdot \frac{d}{dk_1}\left(\frac{1}{k_1} \cdot \frac{d}{dk_1}(F_{11}(k_1,t))\right)\right\} \cdot \sqrt{k_1^2 - k^2} \cdot dk_1 \quad (15)$$

The substituting leads

$$E(k,t) = \frac{2\sqrt{2}}{\Gamma(\sigma)} \cdot a_1^{-2} \cdot (bl^4) \cdot k^3 \cdot \exp\left(-\frac{l^2}{a_1}k^2\right) \cdot U\left(2-\sigma, 3-\sigma, \frac{l^2}{a_1}k^2\right) \quad (16)$$

This is the exact result of the energy spectra based on the Sedov-type solution for two-dimensional turbulence.

In order to obtain the correct asymptotic expansions of the energy spectra in different wave number range, it is important to get the correct asymptotic expansions of the function $U(a,c,z)$.

From the mathematical view of point, we have known that the function $U(a,c,z)$ has very complicate asymptotic forms [14,15,16], which may serve as physical nature of the different asymptotic turbulence spectra behavior, especially for inertial range spectra. By using this conclusion, it is easy to obtain the large scales dynamics of isotropic turbulence in two-dimensional turbulence. Finally, we have

① $0 < \sigma < 2$,

$$E(k,t) = \frac{2}{(2-\sigma)} \cdot \frac{\Gamma(2)}{\Gamma(\sigma)} \cdot a_1^{-\sigma} \cdot (bl^{2\sigma}) \cdot k^{2\sigma-1} \quad (17)$$

② $\sigma = 2$,

$$E(k,t) = 2 \cdot a_1^{-2} \cdot (bl^4) \cdot k^3 \tag{18}$$

③ $\sigma > 2$,

$$E(k,t) = \frac{2\Gamma(2)}{\Gamma(\sigma)} \cdot a_1^{-2} \cdot (bl^4) \cdot k^3 \tag{19}$$

The main results retrieved from the present analysis are as follows:

[1] The parameter $\sigma$ serves as the unique classification index of the asymptotic behavior of turbulent spectra in the low wave number range.

[2] The analysis previously conducted relies on the Sedov-type solution, and demonstrates that a self-similar state whose low wave number kinetic energy spectrum is

$$E(k,t) = C_s \cdot I(t) \cdot k^n \tag{20}$$

The corresponding time evolution of the turbulent kinetic energy spectrum is displayed in the table.1.

It is usually assumed that, for small $k$, the energy spectrum for homogenous two-dimensional turbulence scale as $E(k \to 0) \propto Ik^3$. Here, based on new Sedov-type solution, we have examined possible alternatives, depending on the values of $\sigma$. The investigation which has been presented above indicates that the exact statistical theory for isotropic turbulence is tractable based on the new exact solution of Karman-Howarth equation of two-dimensional turbulence. Analytical study of present theory may be useful in understanding the large scale dynamics for two dimensional isotropic turbulence.

The work was supported by the National Natural Science Foundation of China (Grant Nos.10272018, 10572083).

**Table.1 Time evolution exponents in self-similar decay of isotropic turbulence**

|  | spectra | spectra | spectra |
|---|---|---|---|
| $\sigma$ | $\sigma < 2$ | $\sigma = 2$ | $\sigma > 2$ |
| $C_s$ | $C_-$ | $C_0$ | $C_+$ |
| $I$ | $bl^{2\sigma}$ | $bl^4$ | $bl^4$ |
| $n$ | $2\sigma - 1$ | 3 | 3 |

where

$$C_- = \frac{2}{(2-\sigma)} \cdot \frac{\Gamma(2)}{\Gamma(\sigma)} \cdot a_1^{-\sigma}$$

$$C_0 = 2a_1^{-2}$$

$$C_+ = \frac{2\Gamma(2)}{\Gamma(\sigma)} \cdot a_1^{-2}$$